\documentclass[amsmath,amssymb,reprint,twocolumn,superscriptaddress, nobibnotes, prb]{revtex4-1}

\usepackage[utf8]{inputenc}
\usepackage[T1]{fontenc}
\usepackage[english]{babel}
\usepackage{changes}   
\usepackage[colorlinks=true, citecolor=blue, urlcolor=blue, linkcolor=blue]{hyperref} 

\usepackage{amsmath,amssymb,amsfonts}
\usepackage{amstext, mathrsfs, textcomp}
\usepackage{mathptmx}
\usepackage{bigints}  
\usepackage{nicefrac}

\usepackage{multirow}
\usepackage{dcolumn}
\usepackage{bm}

\usepackage{subfigure}
\usepackage{graphicx}
\usepackage{xcolor}
\usepackage{soul}
\usepackage[export]{adjustbox}

\graphicspath{{Figures/}}

\newcommand{\TITLE}{Frugal Effective Models for Nanophotonic Scattering: Optimizing Global Polarizability Matrices for Metasurface Design}

\setstcolor{teal}

\begin{document}
\title{\TITLE}
\author{\firstname{Sofia} \surname{Ponomareva}}
\affiliation{Univ. Toulouse, CNRS, LAAS, Toulouse, France}
\affiliation{Univ. Toulouse, CNRS, CEMES, Toulouse, France}

\author{\firstname{Peter R.} \surname{Wiecha}}
\email[e-mail~: ]{pwiecha@laas.fr}
\affiliation{Univ. Toulouse, CNRS, LAAS, Toulouse, France}

\begin{abstract}
	Accurate nano-photonics simulations of large scale devices like optical metasurfaces require high accuracy reduced models for the device constituents.
	We present an automated framework for the optimization of Global Polarizability Matrix (GPM) models, which represent a complex scatterer as a small set of non-local effective dipoles. 
	Our goal is to find the most frugal model that reproduces a particle's scattering response within a user-defined accuracy. The method iteratively removes redundant dipoles while re-adapting the positions of the remaining ones via gradient based optimization, stopping at the smallest model that still meets the target. Automatic differentiation, combined with an untrained neural network that reparametrizes the dipole positions, helps to place the dipoles at physically intuitive locations. 
	We demonstrate the versatility of this approach across diverse geometries, from two dimensional ridges over simple spheres to complex three-dimensional particles, achieving compression factors of typically two orders of magnitude compared to full-wave simulations, for target accuracies in the order of few percent. 
	We finally demonstrate how accurate, frugal effective models enable large-scale meta-deflector optimization without periodic approximations.
	This robust recipe for constructing frugal effective models paves the way for the rapid simulation of large-scale photonic assemblies, required for example for metasurface design.
\end{abstract}
\maketitle


\section{Introduction}

Accurate numerical simulations of nano-photonic scattering are crucial for manifold applications in photonics, ranging from imaging through complex media over metasurfaces to surface enhanced spectroscopy.\cite{aouaniUltrasensitiveBroadbandProbing2013, popoffImageTransmissionOpaque2010, elsawyNumericalOptimizationMethods2020}
Full-field methods like the finite difference time domain or the finite element method however are computationally expensive and therefore typically restricted to simulations of single particles or periodic structures.\cite{hsuLocalPhaseMethod2017, schneiderBenchmarkingFiveGlobal2019}
Methods capable of capturing a particle's scattering response in a reduced model are therefore essential for the description of complex scattering problems comprising large, non-periodic arrangements of photonic structures.\cite{sersicMagnetoelectricPointScattering2011, solExperimentallyRealizedPhysicalmodelbased2024, capersDesigningDisorderedMultifunctional2022, munDescribingMetaAtomsUsing2020,majorelDeepLearningEnabled2022, raoThreedimensionalConvolutionalNeural2020, repanArtificialNeuralNetworks2021, herkertInfluenceStructuralDisorder2023}

\begin{figure}[t]
	\centering
	\includegraphics[width=\linewidth]{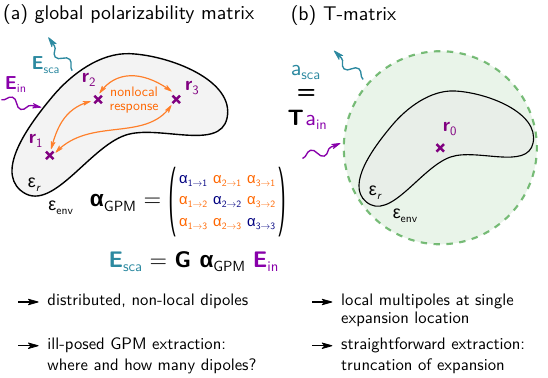}
	\caption{
		Comparison of concepts: Global polarizability matrix (GPM) vs. T-Matrix.
		(a) the GPM links the incoming fields with electric and magnetic dipole moments at $\mathbf{r}_i$. The scattered fields are obtained using superposition of the Green's tensors for field propagation.
		(b) the T-Matrix connects the multipolar expansion coefficients $a_{\text{in}}$ and  $a_{\text{sca}}$ of the incoming, respectively scattered fields, for an expansion around a single position $\mathbf{r}_0$.
	}
	\label{fig:GPM_concept}
\end{figure}

The T-Matrix is a widely used concept to describe the complex optical response of a scattering object.\cite{watermanMatrixFormulationElectromagnetic1965,petersonMatrixElectromagneticScattering1973, mishchenkoTmatrixMethodIts2010, beutelTreamsTmatrixbasedScattering2024}
While in most cases it is very accurate and fast, the T-Matrix also has some limitations.
Most importantly, fields inside the circumscribing sphere around a particle cannot be calculated, which imposes limits for closely packed assemblies, high-aspect ratio structures or scenarios including local light sources like quantum dots.
A possibility to alleviate these limitations is the global polarizability matrix (GPM) concept, recently proposed by Bertrand et al.\cite{bertrandGlobalPolarizabilityMatrix2020}
Instead of a multipole development around a single expansion center, it uses a set of electric and magnetic effective dipoles, distributed inside the original scatterer.
Furthermore, these dipoles are \textit{non-local}, i.e. a field at one dipole location can excite also all other dipoles.
The conceptual differences between GPM and T-Matrix are illustrated in figure~\ref{fig:GPM_concept}.
Thanks to its distributed character, the GPM allows to accurately calculate fields also within the circumscribing sphere.
The method was recently extended to particles that occupy multiple layers in stratified environments.\cite{fuDressedPolarizabilityFramework2026}

While the T-Matrix extraction is a well posed multipole expansion problem and can be done in a straightforward manner,\cite{evlyukhinMultipoleLightScattering2011, alaeeElectromagneticMultipoleExpansion2018, majorelGeneralizingExactMultipole2022, asadovaTmatrixRepresentationOptical2025} the GPM extraction is a more difficult, ill-posed problem.
Two main questions arise: (1) Where to ideally position the effective dipoles? And (2) how many effective dipoles are required?
The question about the positioning has been recently discussed in the context of an expansion around a single location,\cite{ustimenkoOptimalMultipoleCenter2025} and for separate electric and magnetic centers.
\cite{kildishevArtFindingOptimal2025}
In the case of multiple sources, placing them on the topological skeleton was proposed, which can be shown to converge to an asymptotically correct expansion of scattered fields anywhere outside the particle. However for close-to-analytical accuracy this technique constructs models with very large numbers of degrees of freedom.\cite{lamprianidisTranscendingRayleighHypothesis2023}
Determining a practically reasonable number of effective sources for a frugal, yet accurate model remains a challenging task by itself. 
While in the T-Matrix extraction the truncation of the expansion series can be made based on simple error tolerance criteria, with GPM-like methods this is not possible. 
In the latter case the accuracy of a model with a given number of sources is directly connected to the sources' positions. 

In summary, optimizing an effective model based on distributed sources is a complicated, ill-posed problem where the number of expansion terms and the positions of the sources are interconnected and need to be optimized concurrently. 
This is also difficult because it mixes a discrete parameter (the number of sources) with continuous parameters (their positions), and because these parameters are correlated through complex scattering physics.

\begin{figure*}[t]
	\centering
	\includegraphics[width=\linewidth]{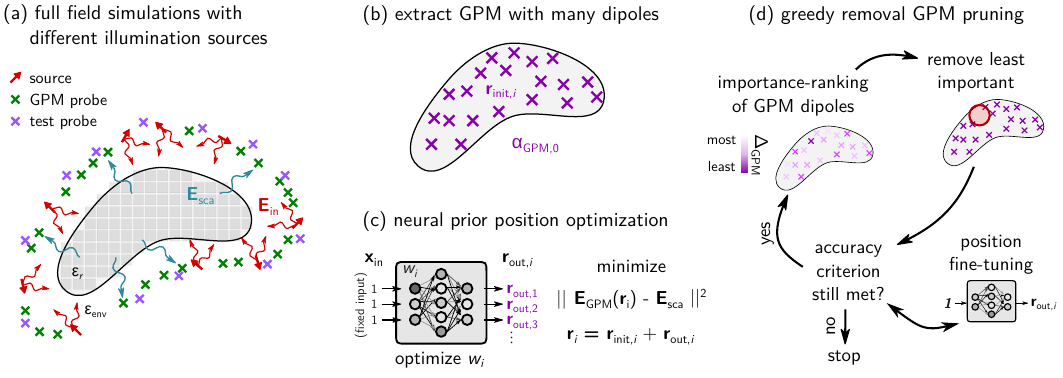}
	\caption{
	GPM optimization procedure.
	(a) Extraction: the GPM is extracted by matching the scattered fields at probe locations for many different illuminations. Evaluate fields also with test configurations, which are used for accuracy evaluation independent of GPM learning. Also use different illuminations for the test cases (not shown).
	(b) An initial GPM is extracted using a large number of effective dipole pairs.
	(c) Initial neural-prior position optimization: the positions $\mathbf{r}_i$ are re-parametrized by an untrained neural network $\mathbf{r}_i = \mathbf{r}_i^{\text{initial}} + \text{NN}(\mathbf{1})_i$, and optimized via autodiff-gradient descent to minimize the field reconstruction loss.
	(d) Greedy removal of effective dipoles: each effective dipole-pair is scored by a leave-one-out re-solve of the GPM (pseudoinverse, cheap) using the reconstruction residual for an importance ranking. The least important dipole-pair is removed. Then, the model is compared to the ``test'' configurations of the full-simulation. If physical accuracy is still better than the user-defined target, repeat. Once the accuracy limit is hit, a fine-tune optimization of the surviving positions is performed to re-center the GPM solution to a more global optimum. Then greedy removal continues.
	Once the accuracy criterion cannot be recovered by the position fine-tune, stop and return the last position fine-tuned GPM that met the criterion.
	}
	\label{fig:gpm_optimization}
\end{figure*}

Recently, automatic differentiation (AD), the key technique for gradient calculation in deep learning,\cite{wengertSimpleAutomaticDerivative1964, baydinAutomaticDifferentiationMachine2018, 
raissiPhysicsinformedNeuralNetworks2019, wenRobustFreeformMetasurface2020, luAgenticFrameworkAutonomous2025} has gained increasing attention in the photonics community as an interesting tool for problems completely unrelated to machine learning.\cite{ganglFullySemiautomatedShape2021, fischbachFrameworkComputeResonances2024}
AD frameworks like PyTorch\cite{paszkePyTorchImperativeStyle2019} or jax\cite{jax2018github} can be used to implement arbitrary calculations like nano-scattering simulations, which then become fully and efficiently differentiable. 
This is highly relevant for gradient based problem solving or physics informed learning.\cite{capersDesigningCollectiveNonlocal2021, huangEigendecompositionfreeInverseDesign2024, asadovaGradientbasedOptimizationScatterer2025, mahlauFDTDXHighPerformanceOpenSource2026, jacksonPyMieDiffDifferentiableMie2025}
For local optimization problems, using \emph{untrained} neural networks has gained attention in the past few years, where the idea is to exploit the inductive bias of the neural network (such as spatial smoothness and low frequency outputs\cite{bostanDeepPhaseDecoder2020, monakhovaUntrainedNetworksCompressive2021}),
which was shown to provide better solutions in a wide range of engineering tasks.\cite{hoyerNeuralReparameterizationImproves2019, Wang2022, zhangTopologyOptimizationImplicit2023}.

Here, we propose to combine gradient optimization through automatically differentiable and GPU accelerated light scattering simulations with a parameter space expansion of the multi-dipole effective model's positional parameters using an untrained neural network (``neural prior'') and a target-accuracy-driven pruning.
Our approach allows finding the source positions for the most frugal GPM within a user-defined accuracy tolerance. 
For pruning, we suggest a greedy elimination of dipole pairs, followed by position finetuning through the neural prior, with a desired accuracy target as stop criterion.
We find that automatic differentiation allows for an efficient and GPU accelerated optimization of the effective dipole locations, and that the neural prior generally improves the position convergence. 
Finally, the accuracy-driven greedy elimination removes unnecessary degrees of freedom of the effective model in a reproducible way.
For visible and infrared sub-wavelength-size particles, we develop a general GPM optimization recipe and provide robust suggestions for its hyperparameters, which leaves the target GPM accuracy as sole user parameter. 
We implement and publish this recipe in the open source autodiff nano-optics simulation toolkit ``TorchGDM''.\cite{ponomarevaTorchGDMGPUacceleratedPython2025}

\section{GPM optimization}\label{sec:GPM_optimization}

To find the optimal number of effective GPM dipole pairs as well as their optimal positions, we propose following procedure consisting of three main steps:
The first step is the extraction of a large GPM, serving as a starting point.
This is followed by the optimization of the GPM dipole positions using a ``neural prior'' reparametrization of the positions through an untrained neural network.
This step regularizes the optimization trajectories through inductive biases of neural networks.\cite{hoyerNeuralReparameterizationImproves2019, joglekarDMFTONNDirectMeshfree2023, zhangTopologyOptimizationImplicit2023, marzbanInverseDesignNanophotonics2026, Wang2022, huangDimensionExpansionSimulationefficient2026}
Finally, a target-accuracy-driven greedy elimination of redundant dipole pairs, followed by position fine-tuning, reduces the number of effective dipoles to the required minimum for a user-defined accuracy tolerance.

\subsection{Extraction of the initial, large GPM}

We start the process by extracting a large GPM, using an arbitrary number $N$ of effective dipoles at predefined positions.
We chose $N=25$ locations $\mathbf{r}_i^{\text{initial}}$ inside the target scatterer volume, distributed based on a simple clustering of the particle volume. At each position of a cluster centroid we place an effective dipole pair.
In our tests with particles several hundred nanometers in size, illuminated by visible to near-infrared light, $N=25$ is a large enough number, well above the tested accuracy criteria. 
Should the initial number of GPM dipole-pairs be not sufficient to reach the target accuracy, our implementation restarts the algorithm with a larger number of dipole pairs (increase by a factor of 1.5). If necessary, $N$ is repeatedly increased until the accuracy criterion is initially met. 

The initial GPM extraction is carried out as proposed by Bertrand et al.:\cite{bertrandGlobalPolarizabilityMatrix2020}
\begin{itemize}
	\item The scattered fields at specified probe locations around the scatterer are computed from full-wave simulations using a large number of different illuminations, as illustrated in figure~\ref{fig:gpm_optimization}a. We use a combination of randomly placed dipole sources and plane waves.\cite{ponomarevaTorchGDMGPUacceleratedPython2025}
	\item The full-field scattering simulations are done using the torchGDM implementation of the Green's Dyadic Method (GDM), a frequency domain volume integral technique.\cite{ponomarevaTorchGDMGPUacceleratedPython2025, girardFieldsNanostructures2005}
	But any other simulation method could be used as well, under the condition that the illumination fields at the GPM dipole positions can be re-created during the extraction and optimization processes.
	\item The GPM parameters are then extracted by solving two inverse problems: The first inverse problem consists in determining the electric and magnetic dipole moments at the GPMs' effective dipole positions that optimally reproduces the scattered fields for each illumination.
	\item The second inverse problem consists in determining the GPM matrix that best reproduces these dipole moments for all given illuminations.
	Both problems are solved using Moore-Penrose pseudoinverse (computed with singular value decomposition, SVD).
\end{itemize}
The hyperparameters which we use for the extraction process (if not otherwise noted in the following), are indicated in table~\ref{tab:hyperparams}. 
Note that we assessed the number of probe locations required to avoid overfitting (see supporting information). 
These tests indicate that around 5 times the number of GPM dipoles is sufficient.
To make sure to avoid any overfitting problems, we use a significantly larger number of probe locations (we use 1500 positions).

\subsection{Optimization of GPM dipole positions using a neural prior}

Starting from this large GPM and using the reference scattered fields, we optimize the initial GPM dipole positions.

To do so, we iteratively recalculate the GPM via SVD. We then use automatic differentiation through the GPM extraction process, to update the effective dipole positions via gradient based minimization of a field reconstruction loss
\begin{equation}\label{eq:loss_reconstruct}
	L_{\text{reconstruct}} = ||E_{\text{GPM}} - E_{\text{sca}} ||^{2} \, ,
\end{equation}
for which we use the mean squared error between the scattered fields from the GPM ($E_{\text{GPM}}$) and from the full simulations ($E_{\text{sca}}$). 
After each coordinate update, the GPM is re-extracted for the new positions.

To render the optimization more robust against local optima, we reparametrize the dipole positions using an \textit{untrained neural network} (see figure~\ref{fig:gpm_optimization}c):
Instead of updating the positions directly, we optimize the weights of a neural network (NN). The NN takes a constant input vector (we simply fix this to ones), its outputs are used as offsets to the initial dipole coordinates:
\begin{equation}
	\mathbf{r}_i = \mathbf{r}_{\text{init},i} + \text{NN}(\mathbf{1})_i \, .
\end{equation}
Specifically, we use a simple feedforward architecture with 2 hidden layers of 256 neurons each, LeakyReLU activation functions inside the network, and linear activations at the outputs.
We initialize the network biases such that all network outputs are zero at the beginning of the optimization.

\begin{figure}
	\centering
	\includegraphics[width=0.95\linewidth]{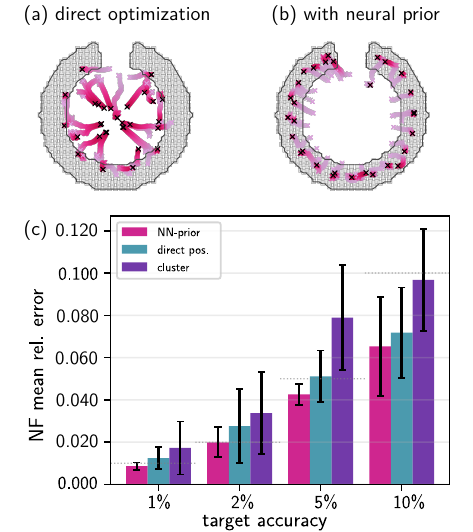}
	\caption{
		GPM dipole position optimization of a silicon split-ring with $60$\,nm height, outer radius of $180$\,nm, inner radius of $120$\,nm and a gap of $\theta_{\text{gap}}=0.5$\,rad, placed in vacuum, illuminated at $\lambda_0=850$\,nm.
		The initial positions of the effective dipoles are chosen very badly on purpose, on a semi-sphere outside of the structure, to illustrate the effect of the neural prior.
		(a) Direct gradient-based position optimization.
		(b) Position optimization with neural prior.
		Both cases use the same random initialization positions.
		(c) Comparison of accuracy with different GPM extraction methods: 
		Optimization with neural prior (purple), direct optimization of the GPM dipole positions (cyan) 
		and without optimization, using clustering and cluster centroids for the GPM dipole locations (violet). 
		Each bar is the average of 9 different structures (see table~\ref{tab:gallery_geometries}). 
		The number of dipoles is determined by optimization to the target accuracy, the three methods are then compared using the same initial set of GPM dipoles. 
		The error bars indicate the standard deviation over all considered structures.
	}
	\label{fig:neural-prior-result}
\end{figure}

On other problems, such ``neural prior'' reparametrization has been found to regularize the search trajectory and to help avoiding local minima.\cite{hoyerNeuralReparameterizationImproves2019, joglekarDMFTONNDirectMeshfree2023, zhangTopologyOptimizationImplicit2023}
On our problem of GPM dipole-pair position optimization, we find that it indeed typically leads to more physically plausible dipole distributions, especially in cases with an unfavorable initial choice for the dipole positions.
We illustrate this first anecdotically by the example of a split-ring particle. With the neural prior, the dipole locations converge toward the structural core, even with a very poor initial choice of GPM positions (Fig.~\ref{fig:neural-prior-result}b), whereas direct optimization of the same initial configuration yields scattered, clustered positions that do not align with the structural features (Fig.~\ref{fig:neural-prior-result}a)
As another illustrative example, we find in the cases of spheres that the direct optimization tends to modify an initial, uniform position distribution only weakly, indicating a local optimum. 
The neural prior reparametrization on the other hand consistently lets all dipoles converge towards the center of spherical particles (see supporting information).

We also performed a systematic comparison between the NN prior, direct positional gradient optimization and a non-optimized position distribution based on clustering. 
We ran both optimizations on 9 different geometries (see table~\ref{tab:gallery_geometries}). 
The different GPMs of each structure have the same, fixed number of dipole pairs.
Their initial positions are the cluster centroids from a clustering algorithm. The statistics of the GPM accuracies are shown in figure~\ref{fig:neural-prior-result}c, where the non-optimized clustering-based GPM (violet bars) is compared to direct optimization (cyan bars), and to the NN prior (purple bars). 
The results indicate that, from an identical starting point, the NN prior leads generally to a better set of positions.

\begin{figure}
	\centering
	\includegraphics[width=0.95\linewidth]{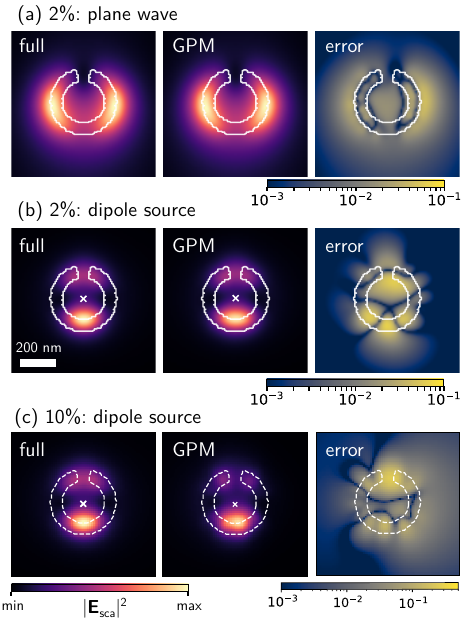}
	\caption{
		GPM nearfield comparison for the split ring structure.
		Scattered nearfield maps, $75$\,nm above the structure, 
		of full simulation vs GPM, $\lambda_0=550\,$nm. (a)-(b) are based on a 2\% mean accuracy target (14 dipole pairs), (c) is based on a 10\% accuracy target (9 dipole pairs). 
		For the shown examples, extraction probes are located at 35\,nm distance from the structure's surface (test probes at 40\,nm).
		(a) normal incidence plane wave illumination, 
		linear polarization towards the split-ring opening, 
		with a peak near-field error of around 4\%, 
		(b) illumination by a dipole light source oriented out-of-plane, located at the center of the split ring (white cross marker),
		In this case the GPM has a peak nearfield error of around 6\%.
		(c) illumination by the same dipole light source, less accurate model gives a peak nearfield error of around 25\%.
	}
	\label{fig:gpm_nearfield}
\end{figure}

\subsection{Target-accuracy-driven greedy elimination with interleaved fine-tuning}

In the third step, we reduce the model complexity to the minimum required for a user-defined accuracy.
The central idea is to let the desired accuracy directly control how many dipole pairs are kept.

\begin{figure*}[t!]
	\centering
	\includegraphics[width=\linewidth]{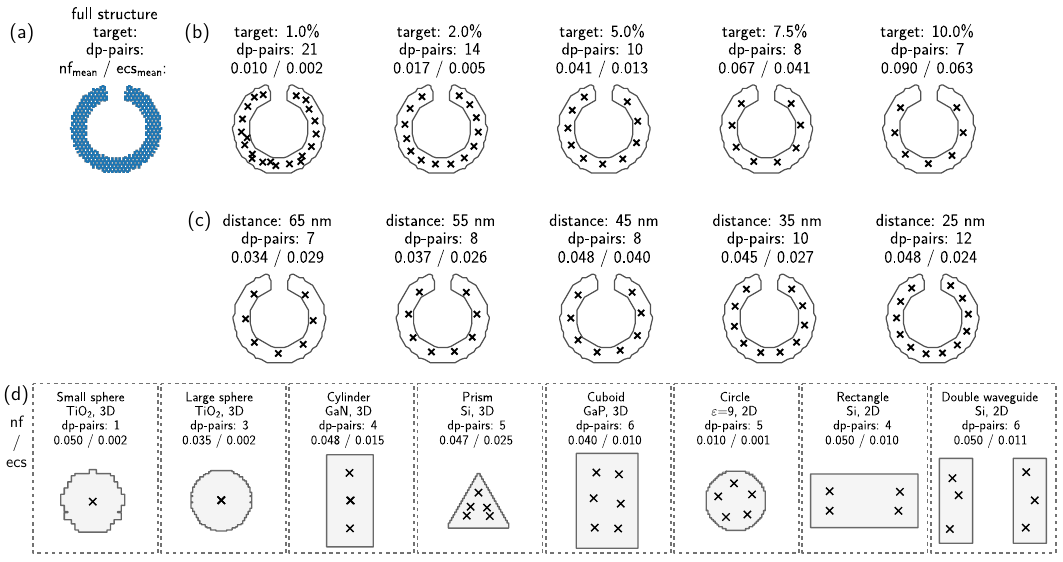}
	\caption{
		GPM optimization example gallery.
		(a) full discretized structure. The silicon split ring has a height of $60$\,nm, an outer radius of $180$\,nm, an inner radius of $120$\,nm and a gap opening of $\theta_{\text{gap}}=0.5$\,rad. It is placed in vacuum and illuminated at $\lambda_0=850$\,nm.
		(b) optimized GPMs for decreasing target accuracy tolerances $\epsilon$ ($1\%$, $2\%$, $5\%$, $7.5\%$, $10\%$). The extraction probe distance is fixed at 50\,nm, the local illumination distance at $80\,$nm. The number of dipole pairs $N$ and the achieved mean relative near-field error and the extinction cross section error are indicated in the title of each panel.
		(c) Series of decreasing GPM extraction probe distances at a fixed accuracy tolerance of $\epsilon = 5\,$\% for the same split-ring structure as in (b).
		(d) Optimized GPMs for various 3D and 2D geometries at a fixed accuracy tolerance of $\epsilon = 5\,$\% (see Table~\ref{tab:gallery_geometries} for geometry details).
	}
	\label{fig:target_series}
\end{figure*}

\textbf{Accuracy target.}
We measure the model accuracy as the mean relative near-field error with respect to the incident field, evaluated on a set of test probe positions and incidences that were \emph{not} used for the GPM extraction.
We randomly generate 500 positions at a fixed distance to the particle surface, using the particle surface normals. 
To ensure test conditions different from the extraction reconstruction loss, we place these positions 5nm farther from the surface than the GPM extraction probes. 
As test illuminations in this accuracy assessment, we use several plane waves and several dipole illumination sources, all at different incident angles, polarization or positions, than the illuminations for the GPM extraction.

The user specifies a target tolerance $\epsilon$ (for instance $\epsilon=0.01$ for a 1\% mean near-field error).
To avoid repeating the expensive full-wave simulation during the pruning, we run it once and cache the reference fields; each candidate GPM is then compared against this cache, requiring only running the cheap GPM simulation.

\begin{table}[t]
	\centering
	\caption{Default hyperparameters of the GPM optimization recipe
		(Section~\ref{sec:GPM_optimization}), unless stated otherwise.
		Distances are referenced to the particle surface.}
	\label{tab:hyperparams}
	\begin{tabular}{@{}ll@{}}
		\hline\hline
		Quantity & Default value \\
		\hline
		Initial dipole pairs $N_0$ & 25 \\
		Extraction probe distance  & 50\,nm \\
		Extraction probe count     & $1500$ ($60\,N_0$ for $N_0=25$) \\
		Test probe distance        & 55\,nm \\
		Test probe count           & $500$ \\
		Probe placement            & random on an offset surface \\
		Plane-wave illuminations   & 10 (5 angles, s/p polarization) \\
		Dipole illuminations       & 60 random positions, \\
		                           & random emitter orientation \\
		Dipole source distance     & 80\,nm \\
		Dipole source placement    & random on an offset surface \\
		Optimizer learning rate    & $5\times10^{-4}$ (same for 2D and 3D) \\
		Optimizer iterations       & optimization stops  \\
		                           & when the loss reaches a plateau \\
		SVD cutoff (pinv)          & PyTorch default: $\max(m,n)\,\varepsilon_{\mathrm{mach}}$ \\
		                           & ($\varepsilon_{\mathrm{mach}}=10^{-7}$, float32), \\
		                           & i.e.\ $\mathcal{O}(10^{-5})$ \\
		Typical runtime            & 5--10\,min on a 6-core Ryzen 3 CPU \\
		                           & (up to 20\,min for larger $N_0$) \\
		\hline\hline
	\end{tabular}
\end{table}

\textbf{Greedy backward elimination.}
Starting from the initial, large GPM model, we iteratively remove the dipole pair whose removal least degrades the model.
To rank the dipoles, we perform a leave-one-out test: for each removed dipole pair, we re-solve the remaining GPM on the reduced set of dipoles via a pseudoinverse. 
This is a single, computationally inexpensive solve per dipole-pair.
We score the candidate by the reconstruction residual $L_{\text{reconstruct}}$ on the extraction probes, which serves as a fast proxy for the physical accuracy.
The dipole pair with the lowest score is removed. 
After removal, we perform a cheap GPM simulation and compare the fields to the pre-calculated test-fields: If the target accuracy is still met, we repeat the dipole-pair removal.
This process is illustrated in figure~\ref{fig:gpm_optimization}d.

\begin{table*}[!b]
	\centering
	\caption{Geometry properties of the structures used in the
		target-accuracy GPM series.  $N_{\mathrm{mesh}}$ is the number of
		volume-discretization mesh cells and $\lambda$ the driving
		wavelength; all structures are embedded in a homogeneous background
		($\varepsilon_{\mathrm{env}} = 1$).}
	\label{tab:gallery_geometries}
	\begin{tabular*}{\linewidth}{@{\extracolsep{\fill}} l c l l r r}
		\hline\hline
		Geometry & Dim. & Material & Size parameters (nm) & $\lambda_0$ (nm) & $N_{\mathrm{mesh}}$ \\
		\hline
		Split ring              & 3D & Si                       & $h=60$, $r_{\mathrm{out}}=180$, $r_{\mathrm{in}}=120$, $\theta_{\mathrm{gap}}=0.5$\,rad & 850 & 1320 \\
		Small sphere            & 3D & TiO$_2$                  & $r = 80$                                                            & 550 & 425  \\
		Large sphere            & 3D & TiO$_2$                  & $r = 160$                                                           & 550 & 3299 \\
		Cylinder                & 3D & GaN                      & $r = 120$, $h = 500$                                                & 550 & 2028 \\
		Trigonal prism          & 3D & Si                       & edge $= 300$, $h = 140$                                             & 550 & 2090 \\
		Cuboid                  & 3D & GaP                      & $300 \times 200 \times 140$                                         & 550 & 2340 \\
		Disc                    & 2D & $\varepsilon = 9$ & $r = 250$                                                           & 550 & 885  \\
		Rectangle               & 2D & Si                       & $400 \times 200$                                                    & 550 & 338  \\
		Double rectangle waveguide & 2D & Si                    & $135 \times 350$ each, sep.\ $300$                                  & 550 & 414  \\
		\hline\hline
	\end{tabular*}%
\end{table*}

\textbf{Stop criterion.}
When a removal breaks the target accuracy, a gradient based ``fine-tune'' optimization of the remaining GPM dipole-pair positions is attempted to \emph{recover} the model.
If the recovered model again meets the accuracy target, the greedy removal continues.
The procedure stops at the smallest number of dipole pairs $N$ for which the position fine-tuned model still meets the target tolerance $\epsilon$.
The result is the minimal-$N$ GPM at the user-specified accuracy $\epsilon$.

\textbf{Fine-tuning frequency.}
The main runtime cost is the position optimization at the initialization and during intermediate position fine-tunings.
We found that the optimal strategy is to perform fine-tuning only when the model's accuracy becomes insufficient during the greedy removal.
Each position optimization runs until the loss does not improve any further (early stopping; we found that this occurs typically after around 100 iterations), but for no more than 500 iterations.
We tested different configurations for the position optimization (see supporting information).
We found that the ``single-pass'' strategy, where position fine-tuning optimizations run only once the accuracy target is no longer met, is typically the fastest approach. 
It requires less fine-tuning runs, while the GPM quality is usually almost identical to pruning with more frequent fine-tuning steps.

\textbf{Typical runtimes.}
On a 6-core Ryzen 3 AMD processor runtimes are in the order of 5-20 minutes, mainly depending on the number of successful final fine-tuning recoveries.

\textbf{Robust hyperparmaeter choice.}
Table~\ref{tab:hyperparams} summarizes the default set of hyperparameters used for the GPM extractions shown in this work. 
We found these parameters to provide robust optimization for scatterers in the visible / near-infrared, and of sizes not larger than the wavelength in the host environment.
Our public implementation of the workflow in torchgdm uses the same default hyperparameter choice.\cite{ponomarevaTorchGDMGPUacceleratedPython2025}

\section{Results and discussion}\label{sec:results}

\subsection{Near field fidelity example}\label{sec:nearfield_example}

We first demonstrate the nearfield reconstruction of a GPM model compared to the full simulation. This is illustrated in figure~\ref{fig:gpm_nearfield} by the example of a silicon split ring (see also Fig.~\ref{fig:target_series}a and table~\ref{tab:gallery_geometries}).
In figure~\ref{fig:gpm_nearfield}a, the structure is illuminated by a normally incident plane wave ($\lambda_0=550$\,nm), with linear polarization along the symmetry axis. 
In figure~\ref{fig:gpm_nearfield}b, the illumination is a local dipole source in the center of the split-ring, with an out of plane emitter orientation. 
Figures~\ref{fig:gpm_nearfield}a-b use a $2\%$ target accuracy.
Figure~\ref{fig:gpm_nearfield}c is based on a target accuracy of  $10\%$, with the same local point source illumination as used in Fig.~\ref{fig:gpm_nearfield}b.

\subsection{Accuracy-target controlled model complexity}\label{sec:tuning_complexity}

In figure~\ref{fig:target_series}b we demonstrate how the target tolerance $\epsilon$ controls the model complexity by the example of the GPM of the same silicon split ring.
Starting from a GPM based on $25$ randomly positioned dipole pairs, we run the optimization for a series of decreasingly tight target tolerances $\epsilon$ ($1\%$, $2\%$, $5\%$, $7.5\%$, $10\%$).
The results are presented in Fig.~\ref{fig:target_series}, where the full discretization is shown in the very left panel, and the subsequent panels show the optimized minimal-$N$ GPMs for increasing tolerance $\epsilon$, which leads to a decreasing number of dipole pairs.
We observed that with increasing number of GPM dipoles the accuracy of the extinction cross-section generally increases faster than the nearfield accuracy. 
We attribute this to the fast decay of higher multipoles' scattered fields in the far-field region, while their contribution can be strong in the near-field region.
Our results demonstrate that GPMs allow typically compression factors of around two orders of magnitude in model complexity, compared to the full coupled dipole simulation.

\subsection{Model complexity for different extraction probe distances}\label{sec:extraction_probe_distance}

In figure~\ref{fig:target_series}c we analyze on the same split ring structure, how the number of GPM dipoles changes when the target accuracy $\epsilon$ is fixed (here to $5$\%) while we change the distance of the GPM extraction probes with respect to the particle surface. 
For a model capable to reconstruct the scattered fields increasingly close to the particle surface, unsurprisingly also the number of required GPM dipoles increases.

\begin{figure*}[t!]
	\centering
	\includegraphics[width=\linewidth]{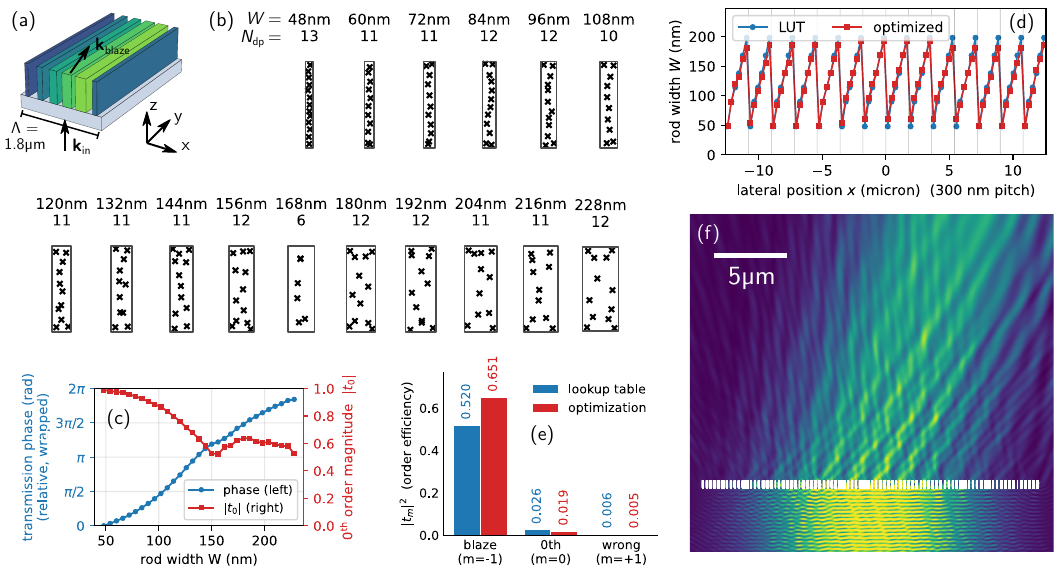}
	\caption{
		Design of a two-dimensional beam-deflecting metagrating from coupled GPM meta-atoms.
		(a) Schematic of the metadeflector unit cell: a finite array of GaN nano-ridges ($H=540\,$nm, pitch $300\,$nm), grouped into $6$-cell blazed supercells that impose a $2\pi$ phase ramp and deflect the normally incident beam into the $m=-1$ blazed order ($\theta\approx 17.8^\circ$).
		(b) Frugal GPM meta-atom library: optimized effective-dipole positions (crosses) for a selection of rod widths, overlaid on the full-rod contours. Each GPM meets the $2\%$ accuracy target with $6$–$14$ dipole pairs.
		(c) Lookup table (LUT): isolated periodic transmission phase (left axis) and $0$th-order magnitude $|t_0|$ (right axis) versus rod width, spanning $\sim 2\pi$ with a smooth amplitude rolloff.
		(d) Rod width versus lateral position across the full $84$-rod device: periodic design taken directly from the lookup table (blue) and holistic aperiodic optimization (red), which chirps the ramp across the aperture.
		(e) Diffraction-order efficiencies $|t_m|^2$ for the blazed ($m=-1$), $0$th ($m=0$) and wrong-side ($m=+1$) orders: lookup table (periodic) design versus the holistic optimization.
		(f) Total near-field intensity $|E|^2$ of the optimized deflector under Gaussian-envelope illumination (incident from below); the beam is deflected into the blazed lobe (mean Poynting flow $+16.8^\circ$).
	}
	\label{fig:metadeflector}
\end{figure*}

\subsection{GPM gallery at a fixed 5\% target}\label{sec:gpm_examples}

In figure~\ref{fig:target_series}d, we apply the same recipe to a selection of different structures, all optimized to the same target accuracy of $\epsilon=5\%$ mean relative near-field error.
The gallery covers three-dimensional particles (small and large TiO$_2$ spheres, a GaN cylinder, a Si prism and a GaP cuboid) as well as two-dimensional geometries with one infinite axis (a circle, a rectangle and a double rectangle structure). The detailed geometrical parameters are given in table~\ref{tab:gallery_geometries}.
For each structure we show the final minimal-$N$ GPM dipole positions within the full geometry's contour. The labels indicate the final number of dipole pairs $N$, as well as mean near-field and extinction cross section errors (Fig.~\ref{fig:target_series}d).
The minimal number of dipole pairs follows the expected trend: a single dipole pair suffices for a small spherical particle, while larger, more complex or more anisotropic shapes (e.g. the cuboid or the double rectangle) require on the order of ten or more dipole pairs to fulfil the accuracy criterion.

\subsection{GPM-based design of a 2D metadeflector array}\label{sec:metadeflector}

As a practically relevant application of frugal GPM models, we consider the design of a metadevice, 
specifically a two-dimensional beam-deflecting metagrating.\cite{elsawyNumericalOptimizationMethods2020, hsuLocalPhaseMethod2017}
The device is a finite array of $84$ GaN nano-ridges ($n\simeq 2.4$, lossless) at $\lambda_0=550\,$nm, arranged in $14$ repetitions of a $6$-cell blazed supercell with a pitch of $300\,$nm (i.e. a supercell period of $\Lambda=1.8\,$\textmu m, see sketch in Fig.~\ref{fig:metadeflector}a).
Each ridge has a fixed height $H=540\,$nm ($\approx 2\lambda_0/n$, the propagation-phase regime), so that the rod width $W$ is the sole design degree of freedom and tunes the local transmitted phase through the effective index.
A supercell of six rods provides six phase steps of roughly $60^\circ$ each, forming a $2\pi$ ramp that directs the normally incident light into the first-order blazed diffraction lobe ($m=-1$) at $\theta_{-1}=\arcsin(-\lambda_0/\Lambda)\approx -17.8^\circ$.

\textbf{GPM-based lookup table.}
We start by building a lookup table (LUT) of GPM models for the meta-atoms: for each of $31$ rod widths ($W=48$\,nm\,$\dots\,228\,$nm, in steps of $6\,$nm) we optimize a frugal GPM with the recipe described in Sec.~\ref{sec:GPM_optimization}, using an accuracy target of $\epsilon = 2\%$.
Starting from $25$ effective dipole pairs, the greedy pruning reduces each meta-atom to a mean of $11.2$ pairs (between $6$ and $14$ dipole pairs at constant accuracy).
A selection of the GPMs is shown in figure~\ref{fig:metadeflector}b.
The isolated, periodic transmission phase of the resulting library sweeps $346^\circ$, i.e. a clean $\sim 2\pi$ design space, with a smooth $0$th-order magnitude $|t_0|$, rolling off from $0.99$ to $0.52$ and no sharp resonance dip (see Fig.~\ref{fig:metadeflector}c).
This pre-computed library is the design space from which every rod in the deflector is drawn.

\textbf{Coupled-GPM array simulation.}
The full metadeflector is assembled as a self-consistently coupled array of all 84 GPM meta-atoms. The frugal character of the GPM models allows to holistically simulate the entire metadeflector at small computational cost, yet retaining all \emph{non-local} inter-cell coupling in accurate simulations at the full device scale.
The distributed character of the GPM models is what makes this tractable: unlike a single-center T-matrix, the effective GPM-dipoles are distributed inside each ridge, so neighbouring meta-atoms can be placed well within their respective circumscribing circles.\cite{bertrandGlobalPolarizabilityMatrix2020, ponomarevaTorchGDMGPUacceleratedPython2025}
Our full $84$-rod device is described by only $\sim 940$ effective dipole pairs, around two orders of magnitude fewer than the $>10^5$ mesh cells that an equivalent full-wave discretization would have. One GPM simulation of the full device takes no more than a few seconds on a normal desktop CPU.

We want to note a technical detail at this point. The edge supercells lack neighbors on one side and therefore scatter differently, which makes the order efficiencies oscillate with the number of periods. To suppress these finite-aperture border effects, the grating is illuminated by a weakly focused Gaussian-envelope plane wave, that gently tapers the edges so the central region sees an effectively periodic environment.

\textbf{Holistic, coupling-aware optimization.}
Because the full GPM metadeflector model is very cheap to calculate, we are capable to optimize every rod of the \emph{full} device independently, allowing the design to break supercell periodicity.
With $84$ cells and $31$ candidate widths, the search space has $31^{84}$ configurations.
As a simple proof of principle, we refine the global design with a simple, fully deterministic greedy polish: in a single pass over the $84$ cells, each rod's width is moved to the neighbouring library width whenever this improves the figure of merit, and only improving moves are kept (algorithmic details, including the convergence of the reported run, are given in the Supporting Information).
We warm-start this optimization from the best periodic design, i.e.\ the supercell of $6$ widths obtained by a coupling-aware coordinate descent on the $6$ supercell meta-atoms, in which every candidate width is evaluated in the full $84$-rod coupled grating (details in the Supporting Information).
The figure of merit is the blazed-order efficiency $|t_{-1}|^2$. We evaluate it in fully coupled simulations using the GPM models, so the inter-cell phase shifts are accounted for automatically: the optimizer selects widths whose \emph{isolated} phases may be non-uniform but become uniform \emph{in the array}.
Finally, we verify the GPM-based optimization solution against a full discretization run using an iterative solver scheme (see supporting information).

We note that in principle, a more sophisticated global search, e.g., a well-adapted simulated-annealing optimization or evolutionary optimization algorithms, may find better designs than the simple greedy polish presented here, but the intention of the shown example is just a feasibility demonstration, showing that the GPM based simulation is fast enough for full-scale optimization. The scope of the present work is not to deliver the best possible optimization of a real world application.

The optimization discovers a genuinely aperiodic assignment, as depicted in figure~\ref{fig:metadeflector}d: every supercell uses a slightly different width ramp, with a wider, higher-phase ramp in the strongly illuminated centre and a compressed ramp at the weakly illuminated edges.
This aperture chirp compensates the Gaussian illumination envelope and is fundamentally inaccessible to a single-supercell design.
It raises the blazed efficiency from $|t_{-1}|^2=0.52$ (best periodic) to $0.65$ (optimized), an improvement of $25\%$. Simultaneously, the optimized design reduces the parasitic $0$th order from $0.026$ to $0.019$ and increases the total transmission from $57\%$ to $69\%$ (Fig.~\ref{fig:metadeflector}e).
We note that the optimization procedure is fully deterministic (no random moves or restarts, convergence details are given in the Supporting Information).
Of the transmitted light, $96\%$ goes into the blazed lobe, and the wrong-side order stays at $\sim 0.5\%$.
The near-field intensity map of the optimized device is shown in figure~\ref{fig:metadeflector}f. The incident beam enters from below and is steadily deflected upwards. The mean Poynting-flow direction above the grating is $+17.2^\circ$, in close agreement with the blazed-order angle of $17.8^\circ$ for the ideal grating.
The total transmission is capped at $\sim 69\%$, with the GaN ridges reflecting $30$--$45\%$ of the incident power. Within a dipolar description, the relevant electric- and magnetic-dipole resonances of the 2D GaN ridge are spectrally separated and cannot be brought into resonance by varying the transverse size alone in the relevant GaN refractive-index regime ($n\simeq2.5$). 
Consequently, the $2\pi$ phase sweep necessarily encounters a predominantly single-dipole Mie resonance rather than a Huygens condition based on spectrally overlapping, phase-matched ED and MD responses. Achieving such a dipolar Huygens regime therefore requires additional geometrical degrees of freedom to independently tune the ED and MD resonances.~\cite{wiechaStronglyDirectionalScattering2017}

\section{Conclusions}

We have presented an automated procedure for optimizing Global Polarizability Matrix models of electromagnetic scatterers, 
based on a set of distributed non-local dipole pairs. By combining differentiable scattering simulations with a neural-prior reparametrization 
of dipole positions, we obtain physically plausible optimized effective models for a fixed number of dipoles. 
Through successive, greedy elimination of the least important dipole pairs, the complexity of the model is subsequently reduced until a user-defined accuracy limit is reached.
We provide an optimized recipe that runs robustly for a wide range of scatterers at visible light / infrared frequencies. 
We also publish an open source implementation of that recipe within the automatic differentiation ready scattering simulation toolkit ``torchgdm''. 
Our results demonstrate that the method is versatile and robust across various geometries, materials and dimensionalities, achieving a compression factor for the model complexity of roughly two orders of magnitude compared to full-wave simulations, for target accuracies of the order of a few percent. 
Our robust and efficient recipe for constructing frugal effective models facilitates the simulation and design of large-scale photonic assemblies such as metasurfaces.
We demonstrate this by full-scale optimization of a two-dimensional beam deflecting metagrating.
The optimization achieves a $25\%$ increase in blazed efficiency compared to the periodic supercell design.
We foresee that 3D, large-scale Huygens metasurface design, where inter meta-atom coupling is an important limiting factor, is precisely the kind of problem for which our frugal-GPM framework is ideally suited.\cite{elsawyNumericalOptimizationMethods2020, wenRobustFreeformMetasurface2020, gigliFundamentalLimitationsHuygens2021}

\begin{acknowledgments}
	We thank Prof. Ulrich Hohenester for fruitful discussions.
	This work was supported by the French Agence Nationale de la Recherche (ANR) under grant ANR-22-CE24-0002 (project NAINOS).
\end{acknowledgments}

\section*{Supporting Information}

The Supporting Information is available free of charge.
In contains additional analysis of GPM optimization convergence, different optimization configurations, and supporting data for the analysis of results (PDF).

\bibliography{2026_ponomareva_optimize_gpms.bbl}
\end{document}